\documentclass[reprint,amssymb,aps,prb]{revtex4-2}
\usepackage{amsmath}
\usepackage{graphicx}
\usepackage{bm}
\usepackage{xcolor}
\begin{document}
\title{Thermal decay of two-spinon bound states in quasi-2D triangular antiferromagnets}

\author{I. L. Pomponio$^{1}$, E. A. Ghioldi$^{2}$, C. J. Gazza$^{1}$, L. O. Manuel$^{1}$, and A. E. Trumper$^{1}$}

\affiliation{$^1$ Instituto de Física Rosario (CONICET) and Facultad de Ciencias Exactas, Ingeniería y Agrimensura, Universidad Nacional de Rosario, 2000 Rosario, Argentina}
\affiliation{$^2$ Department of Physics and Astronomy, University of Tennessee, Knoxville, Tennessee 37996-1200, USA}

\begin{abstract}
We analyze the temperature evolution of the anomalous magnetic spectrum of the spin-1/2 triangular quantum Heisenberg antiferromagnet, which is proximate to a quantum phase transition leading to a spin liquid phase. Recently, its low energy excitations have been identified with two-spinon bound states, well defined in an ample region of the Brillouin zone.  In this work, we compute the thermal magnetic spectrum within a Schwinger boson approach, incorporating Gaussian fluctuations around the saddle-point approximation. In order to account for a finite Néel temperature $T_N$, we incorporate an exchange interaction between triangular layers. As temperature rises, the dispersion relation of the two-spinon bound states, representing single-magnon excitations, remains unchanged but becomes mixed with the thermally activated spinon continuum. Consequently, a crossover occurs at a temperature $T^* \simeq 0.75 T_N$, defining a {\it terminated Goldstone regime} between $T^*$ and $T_N$, where only the magnons close to the Goldstone modes survive as well-defined excitations, up to the Néel temperature.
Our results support the idea that the fractionalization of magnons near a transition to a disordered phase can be extended to more realistic quasi-2D frustrated antiferromagnets.
\end{abstract}

\maketitle

\section{Introduction}
The understanding of the organizing principles~\cite{Laughlin2000} governing frustrated antiferromagnets (AFs) remains a debated topic in condensed matter physics~\cite{Wen2019}. 
The pursuit of quantum spin liquids —highly entangled quantum states without classical analogs and characterized by fractional spinon excitations— has been a central focus since the proposal of the resonant valence bond state for the triangular antiferromagnet~\cite{Anderson1973,Sachdev2008,Normand2009,Powell2011,Savary2017,Zhou2017,broholm2019quantum}. 
This quest has driven both theoretical and experimental efforts, fueled by insights from high-temperature superconductors~\cite{Lee2006} and concepts from the fractional quantum Hall 
effect~\cite{Kalmayer1987}. 
Advancements in numerical and analytical techniques, applied to a variety of frustrated models, have played a relevant role in validating zero-temperature quantum phase diagrams along with their corresponding low-lying magnetic excitations~\cite{Lacroix2011}. 
Additionally, the synthesis of new compounds~\cite{Kim2019} and the refinement of experimental techniques, such as inelastic neutron scattering (INS)~\cite{Tennant2019}, have confirmed the experimental realization of predicted exotic magnetic states of matter. In the current era of quantum materials~\cite{Tokura2017,Keimer2017}, it is crucial to integrate this knowledge to enable theoretical predictions of experiments under diverse conditions in an even more controlled manner.

The paradigmatic model of frustrated quantum AF is the triangular Heisenberg Hamiltonian, whose ground state exhibits a $120^{\circ}$ Néel order, although its proximity to a quantum phase transition (QPT), leading to a quantum spin liquid,  has been well established~\cite{Bernu1992,Capriotti1999,White2007}. The quantum flutuactions, inherent in the AF interactions, are enhanced by magnetic frustration and reduce the local magnetization to $m=0.205$  ($41\%$ of the classical value)~\cite{Capriotti1999}. An indication of its proximity to a QPT is that just a very small second neighbour exchange interaction ($6 \%$ of first neighbour) is enough to destroy the $120^{\circ}$ Néel order~\cite{Zhu15}.

In recent years, the synthesis of triangular antiferromagnets has notably increased, giving rise to numerous candidates for spin liquids~\cite{Li2019,Wen-Jinsheng2019}. However, only few of them are perfectly equilateral,  disorder-free, or faithfully realize a Heisenberg-like Hamiltonian~\cite{Li_2020}.
Instead of directly pursuing highly exotic magnetic states solely characterized by the absence of specific features (such as no ordering down to the lowest temperature or absence of magnon excitations), a more productive strategy has emerged.
Specifically, the study of magnetically ordered or disordered compounds near a QPT has proven to be fruitful. A notable example is the well-studied compound Ba$_3$CoSb$_2$O$_9$ that realizes an effective spin-1/2 triangular antiferromagnet with XXZ model interactions, exhibiting a 120° Néel order below $T_N$ = 3.8K \cite{Doi2004,Susuki2013,Koutroulakis2015,Ma2016,Kamiya2018}. In INS experiments conducted below $T_N$, an unusual spectrum was observed, comprising low-lying energy collective excitations (magnonic-like) coexisting with two dispersive continua at higher energies across the entire Brillouin zone (BZ). The semiclassical theory (large-$S$) \cite{Ma2016,Ito2017,Macdougal20} falls short in explaining this spectrum, leading to speculation that Ba$_3$CoSb$_2$O$_9$ is proximate to a quantum melting point.

Subsequent theoretical studies, including numerical tensor network calculations \cite{Chi2022} and $1/\mathcal{N}$-Schwinger boson (SB) calculations \cite{Ghioldi2022}, support the $XXZ$ model. Furthermore, the SB theory provides a novel interpretation of the spectrum, suggesting that the low-lying energy magnetic excitations consist of two spinon bound states, glued by emergent (Hubbard-Stratonovich) gauge fields. These coexist with two high-energy dispersing continua made of two quasi-free spinons. Namely, at high energies, the confinement length of spinons is greater than the unit lattice constant. In light of these findings, it can be concluded that in the proximity of a quantum melting point, the spin-1 collective excitation of the 120$^{\circ}$ Néel order seems better described by a bound state of two spin-$\frac{1}{2}$ spinons —the quasiparticles of the neighboring spin liquid— than the usual magnonic excitations \cite{Ghioldi2015,Ghioldi2018}.

\begin{figure}[ht]
\vspace*{0.cm}
\includegraphics[width=0.47\textwidth,angle=0]{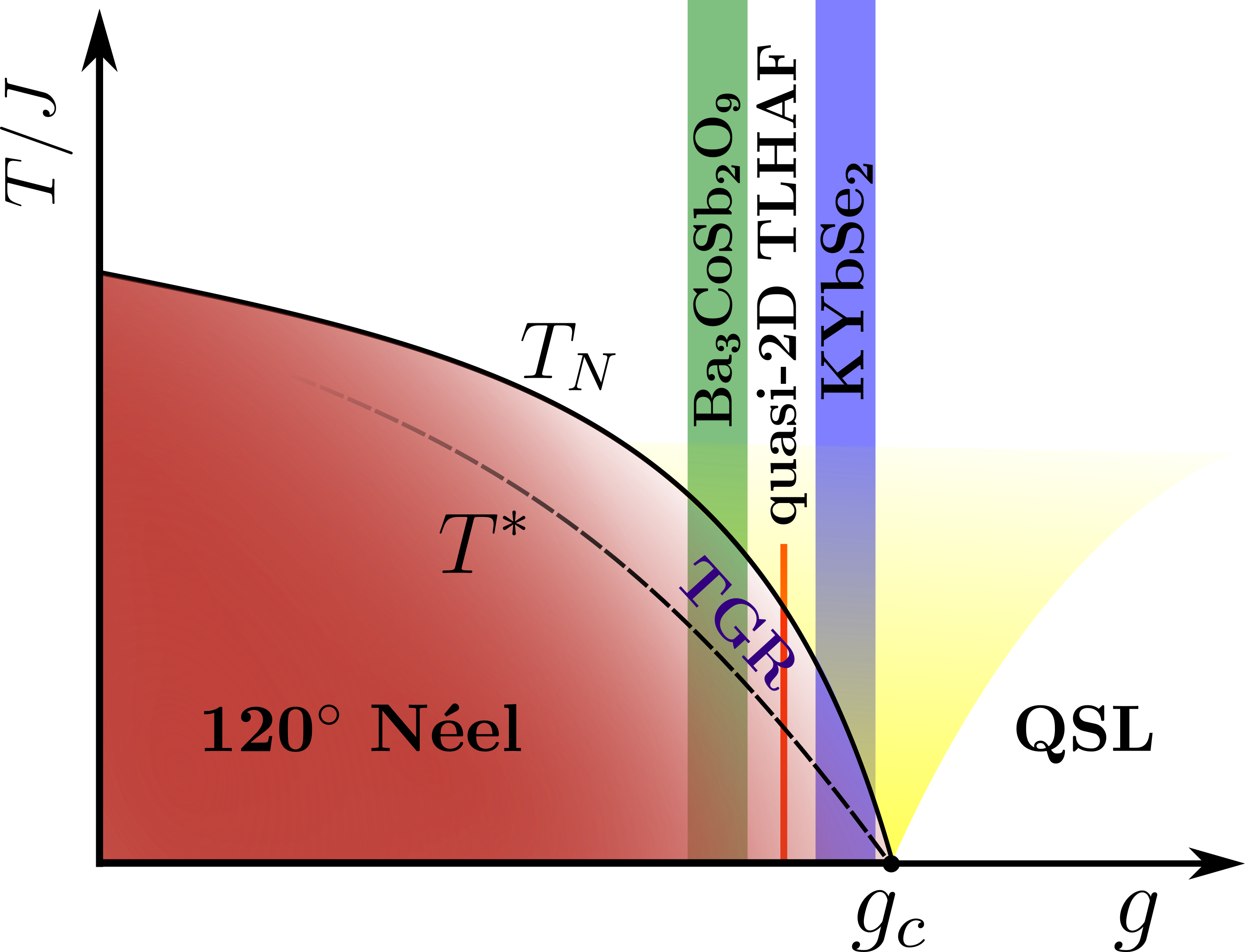}
\caption{
Schematic phase diagram for quasi-2D triangular antiferromagnets~\cite{Chubukov1994, Coldea2003}. The abscissa axis $g$ represents a generic measure of zero-point quantum fluctuations (see text). Solid black line denote the transition from the ordered phase to a disordered one, marked by the N\'eel temperature ($T_N$). The dashed line indicates the spin excitations crossover  around $T^{*}$: from the region with well defined two-spinon bound states across the Brillouin zone to an intermediate one where bound states begin to break up. This region, termed the terminated Goldstone regime (TGR) retains only bound states around the Goldstone modes. The blue and green vertical bands depict the estimated positions of named compounds in the phase diagram. The solid red line represents the location of the model used in our calculations (see Section II). For $g>g_c$, the system undergoes a quantum phase transition to a quantum spin liquid (QSL) phase.}
\label{fig1:diagram}
\end{figure}

More recently~\cite{Scheie24}, the delafossite material KYbSe$_2$, which realizes the $J_1-J_2$ model on the triangular lattice, has been shown to be proximate to a quantum phase transition. In this study, the magnetic spectrum was observed with INS at $T =$ 300 mK, slightly above the Néel temperature $T_N$ = 290 mK. The spectrum displayed an extended continuum with a sharp lower edge that is gapless at the momentum $K$, corresponding to the 120$^\circ$ Néel order expected below $T_N$.
The dynamical structure factor at the magnetic wave vector $K$ exhibits a scaling collapse in $~ \hbar\omega/k_B T$ down to 0.3K, indicative of a second-order quantum phase transition. The agreement with theoretical predictions based on tensor network and $1/\mathcal{N}$ SB calculations suggests that KYbSe$_2$ is even closer to a disordered phase than Ba$_3$CoSb$_2$O$_9$. Based on the zero temperature SB prediction, it is expected that the low-energy edge of the spectrum observed at 0.3K (greater than $T_{N}$) will separate from the continuum as the temperature passes down through the N\'eel temperature, giving rise to collective excitations made of two spinon bound states. However, it remains unclear whether this low-lying energy band will be well-defined across the entire Brillouin zone.
The consistency found between theory and experiments in KYbSe$_2$, along with entanglement witnesses, strongly supports the strategy of investigating the magnetic spectrum of compounds in proximity to a continuous quantum phase transition. In this case, the indications are that KYbSe$_2$ could be proximate to a Z$_2$ spin liquid state.

So far, the theoretical studies mentioned above have primarily been performed at zero temperature, while only a few studies consider finite temperature~\cite{Chen19, Prelovsek18, Gonzalez22}. To provide a more comprehensive description aligned with experimental conditions, the next step is to incorporate the combined effects of both interlayer exchange coupling and temperature. In compounds such as Ba$_3$CoSb$_2$O$_9$ and KYbSe$_2$, although the interlayer interaction is significantly weaker than the intralayer coupling, it plays a crucial role in establishing a finite Néel temperature.

Fig.~\ref{fig1:diagram} schematically depicts the phase diagram of quasi-2D triangular AFs. 
The abscissa axis $g$ represents a generic measure of zero-point quantum fluctuations, with $g_c$ continuously connecting the 120° N\'eel state to a quantum spin liquid state at $T =$ 0. The $g$ parameter can be reduced by including easy plane exchange anisotropy and increased by adding second nearest neighbours interactions, as occurs in Ba$_3$CoSb$_2$O$_9$~\cite{Ma2016} and KYbSe$_2$~\cite{Scheie24}, respectively.  The quasi-2D triangular Heisenberg model is also shown in the magnetic phase diagram for small interlayer exchange interaction.

In this work, we investigate the temperature effect on the magnetic excitation spectra of spin 1/2 quasi-2D triangular AF. Unlike spin wave theory, the  Schwinger boson theory (SBT) relies on magnetic link degrees of freedom that are believed to take better into account the mixing of transverse and longitudinal fluctuations, as soon as the N\'eel temperature is approached. First, we find that at the saddle point (SP) level the SBT recovers qualitatively well the expected (static) behavior of the  $120^{\circ}$ magnetically ordered regime below the Néel temperature. Second, at the $1/\mathcal{N}$ approximation level, we have found an intriguing temperature dependence of the spectrum. Specifically, we identify a low-temperature regime, below a crossover temperature $T^*$, where the two-spinon bound state band is well-defined across an ample region of the Brillouin zone. However, as temperature rises above $T^*$, the gauge field fluctuations fail to form the two-spinon bound states in a significant region of the Brillouin zone. This leads to a differentiation in the magnetic excitation spectrum. Remarkably, for temperatures around $T_N$, only two spinon bound states, centered around the Goldstone modes, remain well-defined. This behavior is similar to what happens in weakly 3D interacting antiferromagnetic chains~\cite{Lake2005}.  In these systems, clear signatures of quantum criticality are observed in the ordered regime (below but close to the N\'eel temperature).  In analogy to the superfluid $^4$He case, where phonon excitations terminate at some momentum and energy by decaying into two roton excitations, we call this antiferromagnetic region --between $T^{*}$ and $T_N$-- {\it terminated Goldstone regime} (TGR) (see Fig.~\ref{fig1:diagram}), although the decay mechanisms are different~\cite{Stone2006}. These finite temperature results strongly support the proposed behavior of frustrated antiferromagnets near a quantum phase transition by Chubukov, Sachdev, and Senthil~\cite{Chubukov1994}, and provide a deeper insight into the fractionalization process~\cite{Chubukov1995}.

In Section II, we briefly developed the $1/\mathcal{N}$ Schwinger boson formalism. Section III includes the interlayer coupling, enabling the study of the Néel temperature. In Section IV, we analyze the effect of temperature on the magnetic excitation spectrum. Section V concludes with final remarks.

\section{Schwinger boson formalism} 

In this section we extend the $1/\mathcal{N}$ SB theory to finite temperatures for quasi two dimensional systems. In order to get a self contained description   we present the main steps of the calculation, emphasizing the temperature dependence of the results. The details of the full calculation can be found in Ref.{\cite{Arovas1988,Auerbach1994,Trumper1997,Ghioldi2018,Zhang2019}}.\\

We study the Heisenberg Hamiltonian in the layered triangular lattice 
\begin{equation}
\hat{H} = \sum_{<ij>}J_{ij} \hat{\textbf{S}}_i \cdot \hat{\textbf{S}}_j, 
\label{heis}
\end{equation}
where $<ij>$ sums over all the nearest-neighbor pairs of spins of the layered triangular lattice. The in-plane exchange interaction is $J$ while there is a no frustrating interlayer exchange interaction $J_\perp$. For simplicity, here we study the isotropic Heisenberg model  with only nearest-neighbor couplings. In what follows, we take $J=1$ as the unit of energy. However, the theory can  be straightforwardly extended to include anisotropic~\cite{Ghioldi2022} and further exchange interactions \cite{Scheie24}. In fact, in the next Section, we will present some Néel temperature estimations for real compounds with more evolved Hamiltonian than the isotropic Heisenberg model.\\

Within the SB formalism the Heisenberg Hamiltonian (\ref{heis}) can be written exactly as 
\begin{equation}
\hat{H} = \sum_{<ij>}J_{ij}  \left(: \hat{B} ^\dagger _{ij} \hat{B} _{ij}: - \hat{A}^\dagger _{ij} \hat{A}_{ij} \right),
\label{BBAA}
\end{equation}
where $ \hat{A}_{ij} = \frac{1}{2}(\hat{b}_{i\uparrow}\hat{b}_{j\downarrow}-\hat{b}_{i\downarrow}\hat{b}_{j\uparrow})$ and $\hat{B}_{ij} = \frac{1}{2}(\hat{b}_{i\uparrow}\hat{b}^\dagger _{j\downarrow}+\hat{b}_{i\downarrow}\hat{b}^\dagger _{j\uparrow})$ are the SU(2) invariant link operators of the theory. $A^{\dagger}_{ij}$ creates singlets between $ij$ sites and $B^{\dagger}_{ij}$ makes them resonate. Furthermore, up to a constant, $:\hat{B} ^\dagger _{ij} \hat{B} _{ij}: \sim (\hat{\textbf{S}}_i+\hat{\textbf{S}}_j)^2$ and $\hat{A}^\dagger _{ij} \hat{A}_{ij} \sim (\hat{\textbf{S}}_i-\hat{\textbf{S}}_j)^2$ which gives a measure of ferromagnetic and antiferromagnetic correlations, respectively. Namely, the particular structure of Hamiltonian (\ref{BBAA}) is ideal to treat Heisenberg models in the presence of magnetic frustration \cite{Ceccatto1993, Flint2009, Mezio2011, Ghioldi2018}.    

The partition function for the model is expressed as the following path integral over the coherent states of the SB's
\begin{equation}
\begin{split}
    \mathcal{Z}[j] = \int \mathcal{D}[\bar{b},b] \mathcal{D}[\lambda] & e ^{ -\int_{0} ^\beta d\tau \left[ \sum_{i\sigma} \bar{b}_{i\sigma} ^\tau \partial _\tau {b}_{i\sigma} + \mathcal{H}(\bar{b},b) + \mathcal{J}_b+ \mathcal{J}_s \right] } \\ \times \  &
    e ^{ -\int_{0} ^\beta d\tau \ i\sum_{i} \lambda_i ^\tau \left( \sum_{\sigma}  \bar{b}_{i\sigma} ^\tau b_{i\sigma}-2S \right)},
\end{split}
\label{generatriz}
\end{equation}
where $\mathcal{H}(\bar{b},b)$ is the Hamiltonian evaluated at site $i$ and time $\tau$ dependent complex eigenvalues of the coherent states, $\mathcal{J}_b = \frac{1}{2}\sum_i h_i ^\mu \textbf{b}_i ^{\tau \dagger}\cdot \sigma ^\mu \cdot\textbf{b}_i ^\tau$ --with $\textbf{b}^{\tau\dagger}_i = (\overline{b}^{\tau}_{i\uparrow}, \overline{b}^{\tau}_{i\downarrow})$-- is the coupling between the spins and an infinitesimal symmetry breaking field $\textbf{h}_i=h(\cos (\textbf{Q}\cdot\textbf{r}_i),\sin (\textbf{Q}\cdot\textbf{r}_i),0)$ that selects the $120^{\circ}$ N\'eel order in the x-y plane [$\textbf{Q}=\left( \frac{2}{3}\pi,\frac{2}{\sqrt{3}}\pi,\pi\right)$] , $\mathcal{J}_s = \frac{1}{2}\sum_i j_i ^{\tau \mu}  \ \textbf{b}_i ^{\tau \dagger } \cdot \sigma^\mu \cdot \textbf{b}_i ^\tau$ is the coupling between the spins and a external magnetic field $j_i ^{\tau \mu}$ used to compute the dynamical spin susceptibility, and $\lambda_i ^\tau$ are the Lagrange multipliers introduced to satisfy the local constraints $b^{\dagger}_{i\uparrow}b_{j\uparrow}+b^{\dagger}_{i\downarrow}b_{j\downarrow}=2S$.

The standard procedure involves the following i) a Hubbard-Stratonovich (HS) transformation to decouple the $A^{\dagger}_{ij}\!A_{ij}$ and $B^{\dagger}_{ij}\!B_{ij}$ terms (quartic in bosons) of $\mathcal{H}(\bar{b},b)$, 
ii) using a local reference quantization axis for the spinors $b^{\tau}_{i\sigma}$ oriented in the direction of the $120^{\circ}$ magnetic order $\textbf{Q}$, iii) integrating out the quadratic bosonic part giving rise to 
\begin{eqnarray}
 \mathcal{Z}_{\rm bos}(\overline W, W, \lambda,j)\!\!&=&\!\!\! \int\! D [\overline b,b] 
 e^{- \vec b^\dagger \cdot \mathcal{G}^{-1}(\overline W, W, \lambda,j) \cdot \vec b}=
 \nonumber \\ 
 &= & \det \left[ \mathcal{G}(\overline W, W, \lambda,j)\right],\nonumber 
\end{eqnarray}
where $\vec b$ is a four component vector containing the variables $b_{i\sigma}^{\tau}$ and $\overline{b}_{i\sigma}^{\tau}$, $\mathcal{G}^{-1}$ is the $4 \times 4$ dynamical matrix (see next subsection), and $W_{ij} ^{\xi \tau}$ (with $\xi = A, \ B$) are the HS auxiliary fields, and finally iv) exponentiate 
$\mathcal{Z}_{\rm bos}(\overline W, W, \lambda,j)$, rendering the total partition function as
\begin{equation}
    \mathcal{Z}[j] = \int \mathcal{D}[\overline{W},W] \mathcal{D}[\lambda]  e ^{ -S_{\rm eff} ( \overline{W},W,\lambda,h,j ) }
    \label{effective}
\end{equation}
with 

\begin{equation*}
{\textstyle S_{\rm eff} ( \overline{W},W,\lambda,h,j) = S_0(\overline{W},W,\lambda) + S_{\rm bos} (\overline{W},W,\lambda,h,j)}    
\end{equation*}
where
\begin{equation}
    S_0(\overline{W},W,\lambda) = \int_0 ^\beta d\tau \big( \sum_{ij,\xi} J_{ij} \overline{W}_{ij} ^{\xi \tau} W_{ij} ^{\xi\tau} -i2S\sum_i \lambda _i ^\tau \big)
\end{equation} 
and
\begin{eqnarray}
 S_{\rm bos}(\overline W, W, \lambda,j)  & \!= \!&- \frac{1}{2} \ln \ \mathcal{Z}_{\rm bos}(\overline W, W, \lambda,j)=   \nonumber \\
                                         &\!=\!& \frac{1}{2} {\rm Tr} \ln \left[ \mathcal{G}^{-1}(\overline W, W, \lambda, j) \right], 
\end{eqnarray}
with the trace, in the last line, taken over space, time, and  boson flavor indices. Equation (\ref{effective}) means that the interacting spin problem has been mapped exactly to a problem of free spin-$\frac{1}{2}$ bosons interacting with fluctuating (space and time) auxiliary fields $W_{ij} ^{A \tau}$, $W_{ij} ^{B \tau}$, and $\lambda_i ^\tau$ which are the gauge fields of the theory \cite{Ghioldi2018}.

 Next, a saddle point expansion is performed and, keeping up to the Gaussian fluctuations of the auxiliary fields, the partition function yields 

\begin{equation}
\mathcal{Z}^{(2)}\![j]\! =\! e^{-S^{\rm sp}_{\rm eff}{(\overline{W}_{\rm sp},{W}_{\rm sp},\lambda_{\rm sp},j)}}\!\times\!\int \! \!D[\bar \phi,\phi]\! \ 
e^{-\Delta \vec \phi_{}^{\dagger}\cdot S_{}^{(2)} \cdot \Delta \vec \phi_{}},
\label{zeta2}
\end{equation}
where the first exponential is the partition function within the SP approximation, and the integral corresponds to the contribution of the 
Gaussian fluctuations of the auxiliary fields $\Delta \vec{\phi}^{\dagger} = \vec{\phi}^{\dagger} - \vec{\phi}^{{\rm sp} \dagger} $  with 
$\vec{\phi}^{\dagger} = \left( \overline W_{i,\delta}^{\xi \tau},\ W_{i,\delta}^{\xi  \tau},\  \lambda_{i}^{\tau} \right)$. $S_{}^{(2)}$ is the fluctuation matrix composed by the second derivatives of the effective action $S_{\rm eff}$ with respect to the auxiliary fields \cite{Ghioldi2018}. It is worth to stress that once the SB theory is extended to a large $\mathcal{N}$ number of flavors, the contribution of Gaussian fluctuations to the dynamical spin susceptibility is of order $1/\mathcal{N}$, so that, we call the present theory $1/\mathcal{N}$ SB theory although we are working within the physical case $\mathcal{N}=2$ \cite{Arovas1988,Auerbach1994,Zhang2019}.

\subsection{Saddle point equations}
The  SP approximation requires  that 
\begin{equation}
     \frac{\partial S_{eff}}{\partial \phi _\alpha} \bigg|_{sp}= \frac{\partial S_{0}}{\partial \phi _\alpha}\bigg|_{sp} + \frac{1}{2} \text{Tr} \left[ \mathcal{G}_{sp} \frac{\partial \mathcal{G}^{-1}}{\partial \phi _\alpha} \bigg|_{sp} \right] = 0.
     \label{speq}
\end{equation}
where $\phi_\alpha$ represents a given HS field (including $\lambda$) at some position and time.
Taking the {\it ansatz} compatible with the magnetic order in the $xy$ plane $\langle A_{\boldsymbol{\delta}}\rangle = - \langle \overline{A}_{\boldsymbol{\delta}}\rangle = iA_{\boldsymbol{\delta}} $ , $\langle B_{\boldsymbol{\delta}}\rangle =  \langle \overline{B}_{\boldsymbol{\delta}}\rangle = B_{\boldsymbol{\delta}} $ ($\boldsymbol{\delta}$ represents the vector that connects sites $i$ and $j$) and $\lambda _{sp} = i\lambda$, the dynamical matrix in the frequency and momentum representation, using the spinor basis $\textbf{b}^{\omega \dagger} _\textbf{k}  = ( \bar{b}^\omega _{\textbf{k}\uparrow},{b}^{-\omega} _{-\textbf{k}\downarrow},\bar{b}^\omega _{\textbf{k}\downarrow},{b}^{-\omega} _{-\textbf{k}\uparrow} ) $, is written as
\begin{widetext}
\begin{equation} \scriptsize
    \mathcal{G}^{-1 }_{sp}(\textbf{k},i\omega) = \begin{pmatrix}
    \left( i\omega+\lambda_{sp}+\gamma ^B _{\textbf{k}+\frac{\textbf{Q}}{2}} \right) e^{-i\omega 0^+} & -\gamma ^A _{\textbf{k}+\frac{\textbf{Q}}{2}}& \frac{h}{2} & 0 \\
    -\gamma ^A _{\textbf{k}+\frac{\textbf{Q}}{2}} & \left(-i\omega+\lambda_{sp}+\gamma ^B _{\textbf{k}+\frac{\textbf{Q}}{2}} \right) e^{i\omega 0^+} & 0 & \frac{h}{2} \\    
      \frac{h}{2} & 0 & \left(i\omega+\lambda_{sp}+\gamma ^B _{-\textbf{k}+\frac{\textbf{Q}}{2}} \right) e^{-i\omega 0^+} & -\gamma ^A _{-\textbf{k}+\frac{\textbf{Q}}{2}} 
    \\    
    0 & \frac{h}{2}& -\gamma ^A _{-\textbf{k}+\frac{\textbf{Q}}{2}} & \left(-i\omega+\lambda_{sp}+\gamma ^B _{-\textbf{k}+\frac{\textbf{Q}}{2}}\right)e^{i\omega 0^+} \end{pmatrix}
    \label{invgreensp}
\end{equation}

\end{widetext}
with functions $\gamma _{\boldsymbol{k}}^{A}\!\!\!=\!\!\! \sum_{\boldsymbol{\delta}>0} J_{\boldsymbol{\delta}} A_{\boldsymbol{\delta}} \sin(\textbf{k}\cdot \boldsymbol{\delta})$ and $\gamma _{\boldsymbol{k}} ^{B} = \sum_{\boldsymbol{\delta}>0} J_{\boldsymbol{\delta}} B_{\boldsymbol{\delta}}
\cos(\textbf{k}\cdot \boldsymbol{\delta})$. 
Convergence factors $e^{\pm i\omega 0^+}$ are necessary to correctly take into account the temporal ordering within the functional integral. 
The inverse of the dynamical matrix at the SP is the spinon propagator $\mathcal{G}_{sp}$ that can be decomposed in term of simple fractions as
\begin{equation}
    \mathcal{G}_{sp}(\textbf{k},i\omega) = \sum_{\sigma \sigma '} \frac{g ^{\sigma \sigma '} \scriptstyle{(\textbf{k})}  }{i\omega + \sigma \varepsilon_\textbf{k} ^{\sigma '}(h)}
    \label{greensp}
\end{equation}
whose poles are the dispersion relations of the free spinons $\varepsilon_\textbf{k} ^{\sigma}(h)$, with $\sigma, \sigma'=\pm$, that can be obtained together with the matrices $g^{\sigma \sigma '}$ after performing a paraunitary diagonalization of the dynamical matrix (See Appendix A). Replacing the expressions of $\mathcal{G}^{-1 }_{sp}$ and $\mathcal{G}_{sp}$ in the SP equation (\ref{speq}), and carrying out the Matsubara sums, we arrive at the self-consistent set of equations
\begin{align}
    A_{\boldsymbol{\delta}} &= \frac{1}{2N_s} \sum_{\textbf{k}} \tilde{A}_\textbf{k} \sin (\textbf{k}+\frac{\textbf{Q}}{2} )\cdot \boldsymbol{\delta}\label{A} \\
        B_{\boldsymbol{\delta}} &= \frac{1}{2N_s} \sum_{\textbf{k}} \tilde{B}_\textbf{k} \cos (\textbf{k}+\frac{\textbf{Q}}{2} )\cdot \boldsymbol{\delta}\label{B} \\
        S &= \frac{1}{2N_s} \sum_{\textbf{k}} \tilde{B}_\textbf{k} \label{const}
\end{align}
where $N_s$ is the total number of sites and
\begin{equation}
\begin{split}
    \tilde{A}_\textbf{k} &= \sum_{\sigma \sigma '} \left( g ^{\sigma \sigma '} _{21} {\scriptstyle{(\textbf{k})}}
    + g ^{\sigma \sigma '} _{43} \scriptstyle{(-\textbf{k})}\right) \left[ 1+n(\sigma \varepsilon _{\textbf{k}} ^{\sigma'}(h))  \right] \\
    \tilde{B}_\textbf{k} &=  \sum_{\sigma \sigma '}  g ^{\sigma \sigma '} _{11} {\scriptstyle{(\textbf{k})}}\ n(\sigma \varepsilon _{\textbf{k}} ^{\sigma'}(h)) 
    + g ^{\sigma \sigma '} _{22} {\scriptstyle (\textbf{k})} \left[1+n(\sigma \varepsilon _{\textbf{k}} ^{\sigma'}(h)) \right].
\end{split}
\label{AB}
\end{equation}
The explicit temperature dependence is present in the Bose occupation numbers $n$.
\subsection{Dynamical spin susceptibility}
The dynamical spin susceptibility is obtained as \cite{Auerbach1994,Ghioldi2018}
\begin{equation}
     \chi _{\mu\nu} (\textbf{q},i\omega) = \lim_{h\rightarrow 0} \lim_{N_s\rightarrow \infty} \frac{\partial ^2 \ln \mathcal{Z}[j] }{\partial j^\mu _{\textbf{q},i\omega} \partial j^\nu _{-\textbf{q},-i\omega}} \bigg|_{j=0},
    \label{suscep}
\end{equation}
where the order of the limits is a key point of the calculation since we are interested in the excitation spectrum of magnetically ordered states proximate to a quantum melting point. In particular, to compute the dynamical spin susceptibility up to Gaussian order --above the SP solution-- $\mathcal{Z}^{(2)}\![j]$ of Eq.~(\ref{zeta2}) is plugged in Eq.~(\ref{suscep}), giving rise to three terms of order $1/\mathcal{N}$. In Refs.~\cite{Ghioldi2018,Zhang2019} we have shown that by only keeping the bubble like diagram the dynamical susceptibility can be written as  
\begin{equation}
    \chi _{\mu\mu} (\textbf{q},i\omega) = \chi^{sp} _{\mu\mu} (\textbf{q},i\omega) + \chi^{fl} _{\mu\mu} (\textbf{q},i\omega),
\end{equation}
being
\begin{equation}
    \chi^{sp} _{\mu\mu} (\textbf{q},i\omega) = \frac{1}{2} \text{Tr} \left[ \mathcal{G}_{sp} \frac{\partial \mathcal{G} ^{-1}}{\partial j^{u} _{\textbf{q},i\omega} } \bigg|_{sp} \mathcal{G}_{sp} \frac{\partial \mathcal{G} ^{-1}}{\partial j^{u} _{-\textbf{q},-i\omega} }\bigg|_{sp} \right]
    \label{chisp}
\end{equation}
and
\begin{equation}
      \chi^{fl} _{\mu\mu} (\textbf{q},i\omega) = \sum_{\alpha_1 \alpha_2} \Lambda ^{\mu} _{\phi_{\alpha_1}} (\textbf{q},i\omega) \mathcal{D}_{\alpha_2 \alpha_1}(\textbf{q},i\omega) \Lambda ^{\mu} _{\phi_{\alpha_2}} (-\textbf{q},-i\omega)
\label{eq:corrections}
\end{equation}
where
\begin{equation}
        \Lambda ^{\mu} _{\phi_{\alpha_i}} (\textbf{q},i\omega) = \frac{1}{2} \text{Tr} \left[ \mathcal{G}_{sp} \frac{\partial \mathcal{G} ^{-1}}{\partial \phi_{\alpha_i} } \bigg|_{sp} \mathcal{G}_{sp} \frac{\partial \mathcal{G} ^{-1}}{\partial j^{u} _{\textbf{q},i\omega} }\bigg|_{sp} \right],
        \label{lvector}
\end{equation}
and $\mathcal{D}_{\alpha_1 \alpha_2}$ is the inverse of the fluctuation matrix $S^{(2)} _{\alpha_1 \alpha_2}=\frac{1}{2}\frac{\partial ^2 S_{eff}}{\partial \phi_{\alpha_1} \partial \phi_{\alpha_2}} \big|_{sp} = \frac{1}{2} \left\{  \frac{\partial ^2 S_{0}}{\partial \phi_{\alpha_1} \partial \phi_{\alpha_2}} - \Pi _{\alpha_1 \alpha_2} \right\}$ with
\begin{equation}
    \Pi _{\alpha_1 \alpha_2}  = \frac{1}{2} \text{Tr} \left[ \mathcal{G}_{sp} \frac{\partial \mathcal{G} ^{-1}}{\partial \phi_{\alpha_1} } \bigg|_{sp} \mathcal{G}_{sp} \frac{\partial \mathcal{G} ^{-1}}{\partial \phi_{\alpha_2} } \bigg|_{sp}  \right].
\label{mpolar}
\end{equation}
After carrying out the Matsubara's frequency sums in (\ref{chisp}), (\ref{lvector}) and (\ref{mpolar}) for the out of plane $zz$ entry of the susceptibility, we arrive at
\begin{widetext}
\begin{equation}
\begin{split}
    \chi _{zz} ^{ sp}(\textbf{q},i\omega) = \frac{1}{8N_s} \sum_{\textbf{k}} \sum_{\sigma \sigma ' \sigma ''} \Bigg\{ &\frac{Tr\left[g^{\sigma \sigma '} {\scriptstyle{(\textbf{k}+\textbf{q})}} u^z  g^{\sigma \sigma ''} {\scriptstyle{(\textbf{k})}} u^z \right]}{\varepsilon ^{\sigma '} _{\textbf{k}+\textbf{q}} - \varepsilon ^{\sigma ''} _{\textbf{k}}+\sigma i \omega} \left( n(\varepsilon ^{\sigma ''} _\textbf{k}) - n(\varepsilon ^{\sigma '} _{\textbf{k}+\textbf{q}}) \right)  - \\ 
    &\frac{Tr\left[g^{\sigma \sigma '} {\scriptstyle{(\textbf{k}+\textbf{q})}} u^z  g^{\bar{\sigma} \sigma ''} {\scriptstyle{(\textbf{k})}}  u^z \right]}{\varepsilon ^{\sigma '} _{\textbf{k}+\textbf{q}} + \varepsilon ^{\sigma ''} _{\textbf{k}}+\sigma i \omega} \left( n(\varepsilon ^{\sigma ''} _\textbf{k}) + n(\varepsilon ^{\sigma '} _{\textbf{k}+\textbf{q}}) + 1 \right) \Bigg\}
\end{split}
\label{eq:chizzsp}
\end{equation}
\begin{equation}
\begin{split}
      \Lambda ^z _{\phi_{\alpha_1}} (\textbf{q},i\omega) = \frac{1}{4N_s} \sum_{\textbf{k}} \sum_{\sigma \sigma ' \sigma ''} \Bigg\{ &\frac{Tr\left[g^{\sigma \sigma '} {\scriptstyle{(\textbf{k}+\textbf{q})}} u^{z }  g^{\sigma \sigma ''} {\scriptstyle{(\textbf{k})}} v^{\textbf{k}, \textbf{k}+\textbf{q}} _{\phi_{\alpha _1}} \right]}{\varepsilon ^{\sigma '} _{\textbf{k}+\textbf{q}} - \varepsilon ^{\sigma ''} _{\textbf{k}}+\sigma i \omega} \left( n(\varepsilon ^{\sigma ''} _\textbf{k}) - n(\varepsilon ^{\sigma '} _{\textbf{k}+\textbf{q}}) \right)  - \\ 
    & \frac{Tr\left[g^{\sigma \sigma '} {\scriptstyle{(\textbf{k}+\textbf{q})}} u^{z}   g ^{\bar{\sigma} \sigma ''} {\scriptstyle{(\textbf{k})}}  v^{\textbf{k}, \textbf{k}+\textbf{q}} _{\phi_{\alpha _1}} \right]}{\varepsilon ^{\sigma '} _{\textbf{k}+\textbf{q}} + \varepsilon ^{\sigma ''} _{\textbf{k}}+\sigma i \omega} \left( n(\varepsilon ^{\sigma ''} _\textbf{k}) + n(\varepsilon ^{\sigma '} _{\textbf{k}+\textbf{q}}) + 1 \right) \Bigg\} 
\end{split}
\end{equation}
\begin{equation}
\begin{split}
      \Pi_{\alpha_1 \alpha_2} (\textbf{q},i\omega) = \frac{1}{2N_s} \sum_{\textbf{k}} \sum_{\sigma \sigma ' \sigma ''} \Bigg\{ &\frac{Tr\left[g^{\sigma \sigma '} {\scriptstyle{(\textbf{k}+\textbf{q})}} v^{ \textbf{k}+\textbf{q},\textbf{k}} _{\phi_{\alpha _1}}  g^{\sigma \sigma ''} {\scriptstyle{(\textbf{k})}}  v^{\textbf{k}, \textbf{k}+\textbf{q}} _{\phi_{\alpha _2}} \right]}{\varepsilon ^{\sigma '} _{\textbf{k}+\textbf{q}} - \varepsilon ^{\sigma ''} _{\textbf{k}}+\sigma i \omega} \left( n(\varepsilon ^{\sigma ''} _\textbf{k}) - n(\varepsilon ^{\sigma '} _{\textbf{k}+\textbf{q}}) \right)  - \\ 
    & \frac{Tr\left[g^{\sigma \sigma '} {\scriptstyle{(\textbf{k}+\textbf{q})}} v^{ \textbf{k}+\textbf{q},\textbf{k}} _{\phi_{\alpha _1}}   g ^{\bar{\sigma} \sigma ''} {\scriptstyle{(\textbf{k})}} v^{\textbf{k}, \textbf{k}+\textbf{q}} _{\phi_{\alpha _2}} \right]}{\varepsilon ^{\sigma '} _{\textbf{k}+\textbf{q}} + \varepsilon ^{\sigma ''} _{\textbf{k}}+\sigma i \omega} \left( n(\varepsilon ^{\sigma ''} _\textbf{k}) + n(\varepsilon ^{\sigma '} _{\textbf{k}+\textbf{q}}) + 1 \right) \Bigg\}
\end{split}
\label{eq:pifl}
\end{equation}
\end{widetext}
where $u_z = diag(1,-1,-1,1)$ come from the external vertices $\frac{\partial \mathcal{G} ^{-1}}{\partial j^{u} _{\textbf{q},i\omega} }$  and $v^{\textbf{k}, \textbf{k}'} _{\phi_{\alpha}}$ come from the internal vertices $\frac{\partial \mathcal{G} ^{-1}}{\partial \phi_{\alpha} }$. Similar expressions can be obtained for the in-plane ($xx$ and $yy$) entries of the susceptibility.

Finally, by means of the fluctuation-dissipation theorem, we can obtain the dynamical spin structure factor by performing the analytic continuation $i\omega \rightarrow \omega +i\eta$ in the dynamical spin susceptibility
\begin{equation}
    S^{\mu \mu} (\textbf{q},\omega) =  \left(\frac{-1}{\pi}\right) \frac{1}{1-e^{-\beta \omega }} \text{Im} \big[ \chi ^{\mu \mu} (\textbf{q},\omega +i\eta) \big],
    \label{eq:dssf}
\end{equation}
where $\eta$ is the artificial broadening. 

\section{Magnetically ordered regime} 

In this section we focus on the ability of the SBT to describe the finite temperature behavior as a function of $J_\perp$. In particular, in the ordered regime, we find that the SP solution recovers qualitatively well the expected (static) behavior of quasi-2D triangular AFs. 

In accordance with the Mermin-Wagner theorem \cite{Arovas1988,Auerbach1994}, the onset of long-range magnetic order in a triangular Heisenberg antiferromagnet --with continuous symmetry and short range exchange interactions-- can only occur at zero temperature. Correspondingly, by mapping the 2D Heisenberg antiferromagnet to the effective non linear sigma model (NLSM), it is shown that there is a crossover from a high temperature paramagnetic regime to a low temperature classical renormalized regime where the finite temperature spin correlation functions increases exponentially as  the temperature goes down to zero \cite{Azaria93}. The SBT reproduces quite well this low temperature behavior of the spin correlations at the saddle point level \cite{Yoshioka1991,Chubukov1994}. In particular, it recovers the low temperature $120^{\circ}$ N\'eel correlation functions predicted by NLSM, while at zero temperature the condensation of the SBs signals the occurrence of a finite magnetization. At higher temperatures, however, the SP approximation overestimates the entropy --due to violation of the local constraint-- so neighboring spins become perfectly uncorrelated at certain temperature \cite{Zhang2002}. This unphysical result has been studied in the context of large $\mathcal{N}$ theory along with the possibility of being corrected for finite $\mathcal{N}$ \cite{Tchernyshyov2002}. 

On the other hand, any finite interlayer exchange $J_\perp$ gives rise to long range magnetic order below a finite Néel temperature. The very definition of a quasi-2D antiferromagnet implies that $J_\perp$ is much smaller than the in-plane exchange interaction $J$ --two or more orders of magnitude smaller--, making it difficult to model the compounds in order to extract $J_\perp$ values from the measured Néel temperatures. 

\begin{figure}[t]
\vspace*{0.cm}
\includegraphics*[width=0.5\textwidth,angle=0]{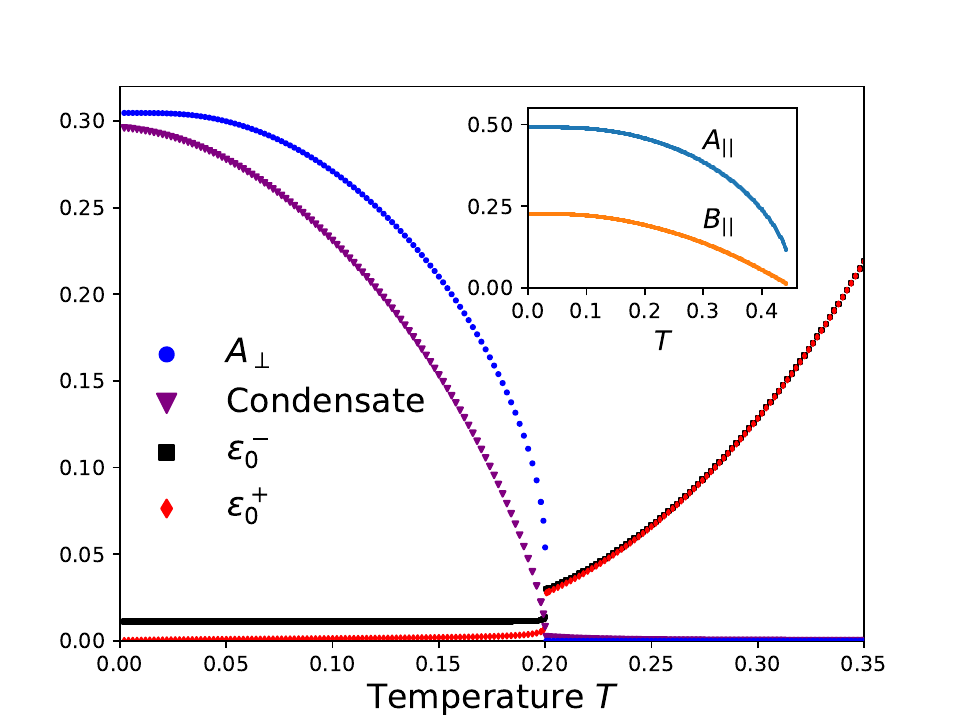}
\caption{$A_\perp$ (blue circles), condensate (purple triangles), 
$\sigma\!=\!+$ (red diamonds) and $\sigma\!=\!-$ (black squares) spinon dispersion relation gaps as a function of the temperature for $N_s= 144\times144\times8$ sites and $J_{\perp}=0.01J$. Inset: Temperature dependence of the in-plane saddle point parameters $A_{\parallel}$ and $B_{\parallel}$.}
\label{fig:bidimensionalizacion}
\end{figure}

Formally, the magnetic phase transition is derived by studying the behavior of the SB self consistent Eqs. (\ref{A}-\ref{const}) in the thermodynamic limit, namely, first taking $N_s\rightarrow \infty$ and then $h \rightarrow 0$ \cite{Auerbach1994}. In particular, there is a regime where $\varepsilon^{+}_{{\bf 0 }}\rightarrow 0$ and $\varepsilon^{-}_{{\bf 0}}\neq0$ which corresponds to the condensation of the SB $\sigma = +$ at ${\bf q } =0$. This condensation is directly related to the presence of a local magnetization that lies in the $xy$ plane and points in the $x$ direction of a local reference frame that rotates 120$^{\circ}$ from site to site. In practice, this calculations is carried out on finite size systems of $N_s = N_{\perp} \times N_{\parallel} $ sites, where $N_{\perp}$ is the number of layers taken and $N_{\parallel}$ is the number of sites in each triangular lattice layer. Then, a symmetry breaking field $h = 1/N_{\parallel}$ is applied. Since $\varepsilon^{+}_{{\bf 0 }} \sim 1/N_{\parallel}$ and $\varepsilon^{-}_{{\bf 0 }} \sim (1/N_{\parallel})^{\frac{1}{2}}$, the condensate is identified with the ${\bf q}=0$ and $+$ term of the constraint equation (\ref{const}). Fig. \ref{fig:bidimensionalizacion} shows the dependence of this condensate with temperature for a quasi-2D triangular system composed of eight layers --with periodic boundary conditions-- and $J_{\perp}/J = 0.01$. In this work, we have taken $N_{\parallel}=144 \times 144$. This figure also shows the spinon relation dispersion values $\varepsilon^{+}_{\bf 0 }$ and $\varepsilon^{-}_{{\bf 0}}$ where it can be seen that $\varepsilon^{+}_{{\bf 0}}\sim 0$ (gapless) and $\varepsilon^{-}_{{\bf 0}}\neq0$ (gapped) for $T\leq T_N$ (consistent with  $\sigma =+$, $\bm q= \bm 0$ condensate); while for $T > T_N$, a gap is opened in both flavors $\varepsilon^{+}_{{\bf 0}}=\varepsilon^{-}_{{\bf 0}}\neq 0$ which signals no condensation, that is, the absence of magnetic order. In the next section we will see that this behavior of the condensate has a crucial effect on the magnetic spectrum as one approaches to the Néel temperature. 

One way to measure the spin correlation between nearest-neighbor spins along the interlayer  direction is by computing the saddle-point value of the Hubbard-Stratonovich field $A_\perp$, through the relation 
\begin{equation}
\langle \bm S_i \cdot \bm S_{i+c}\rangle = B_\perp^2- A_\perp^2.
\label{eq:ssperp}
\end{equation}
As the interlayer exchange interactions are not frustrated, $B_\perp$ vanishes, signaling the antiparallel spin alignment along the perpendicular direction. The behavior of $A_\perp$ versus temperature is shown in Fig.~\ref{fig:bidimensionalizacion}. There we observe that
$A_\perp$ behaves similarly to the condensate, particularly vanishing at the Néel temperature, so that the fact that $A_\perp \neq 0$ is an evidence of the 3D character of the magnetic order. Namely, if $A_\perp \neq 0$, the system exhibits long-range magnetic order, whereas above $T_N$  the triangular layers are magnetically decoupled ($A_\perp=0$) and the local magnetization goes to zero, in accordance with Mermin-Wagner physics. In other words, the transition between the magnetically ordered and paramagnetic phase occurs due to the effective bidimensionalization of the magnetic state as the temperature increases. So, the vanishing of $A_\perp$ is another good signal of the Néel temperature. 
Notice that as the temperature increases, the spin correlations in all directions are weakened.  One would expect, naively, that $A_\perp$ is of the order of $J_\perp,$ however any finite value of the interlayer coupling is enough to generate strong correlations between triangular layers ($A_{\perp} \simeq 0.3$ at $T=0$). 
 
 The inset of Fig.~\ref{fig:bidimensionalizacion} shows the dependence of the in-plane parameters $A_{\parallel}$ and $B_{\parallel}$ with temperature. Even if the SP solution predicts the above mentioned unphysical transition to a perfectly uncorrelated paramagnet around $T\sim 0.43 J $, when $T\lesssim 0.2J $ the SP solution predicts reasonable results for the magnetically ordered regime of quasi-2D triangular AFs.

 \begin{figure}[t]
\vspace*{0.cm}
\includegraphics*[width=0.5\textwidth,angle=0]{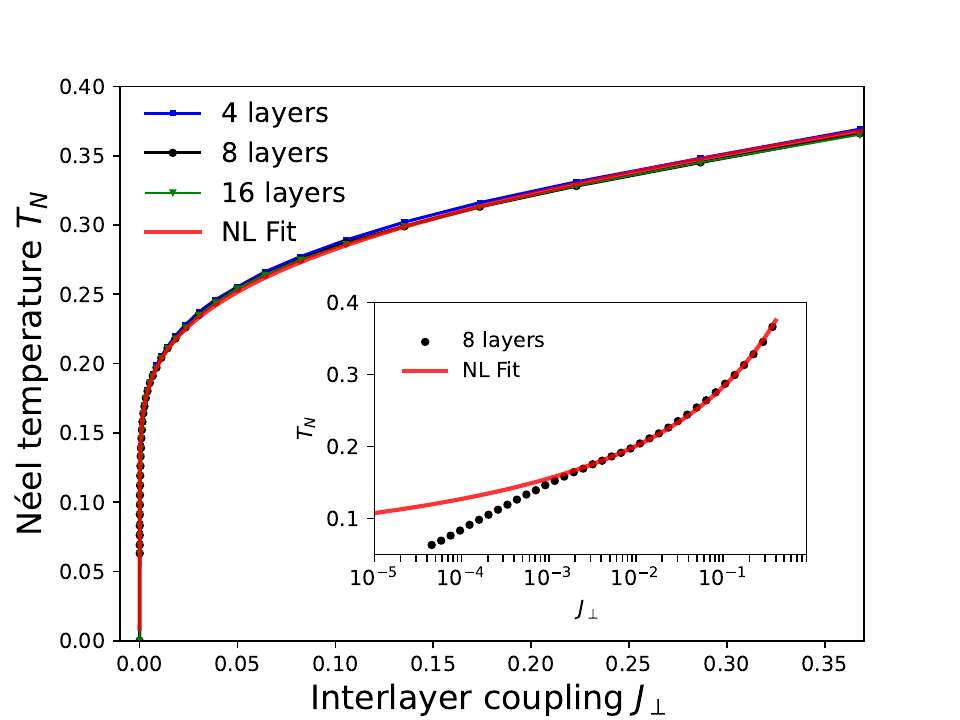}
\caption{Néel temperature as a function of the interlayer coupling $J_\perp$ for $N_{\parallel}=144 \times 144$ and different number of layers $N_\perp$ (with periodic boundary conditions). Inset: The solid red line corresponds to the non-linear fit (\ref{eq:TN}) for the eight layers SB result.}
\label{fig:neelvsjperp}
\end{figure}

Fig.~\ref{fig:neelvsjperp} shows the Néel temperature as a function of $J_{\perp}$ and the number of triangular layers with periodic boundary conditions. For very low  $J_{\perp}$ it is observed that $T_N$ remains practically independent of the number of layers for $N_{\perp} \geq 4$, notoriously simplifying the calculation of $T_N$ for quasi-2D systems.  It should be pointed out that linear spin wave theory (LSWT) applied to the quasi-2D triangular AF predicts larger values of the N\'eel temperatures than the SBT ones~\cite{Du02}. For instance, for  $J_{\perp}/J=0.25$, $T^{LSWT}_N /J\sim 0.8$ while $T^{SBT}_N/J \sim 0.32$. This clearly indicates that LSWT strongly overestimates the finite temperature magnetization with respect to SBT. These values are consistent with the expected behavior near $T_N$, where the transverse and longitudinal fluctuations become of the same order. This feature is surely better captured by the link fields $A_{ij}$ and $B_{ij}$ of the SBT than the local description of LSWT. 
In addition, the SB results are compared with a non linear fit corresponding to the empirical formula 
\begin{equation}
T_N(J_\perp) = \frac{4\pi \rho _s}{b-\ln (J_\perp)}
\label{eq:TN}
\end{equation}
based on the random phase approximation combined with quantum Monte Carlo simulations performed in unfrustrated Heisenberg AFs \cite{Yasuda2005}. Here, the best fit is given by $\rho _s=0.12675$, $b=3.33257$. Notably, the SB results follows quite well the empirical formula down to $J_{\perp} \sim 10^{-3}$ (see inset of Fig.~\ref{fig:neelvsjperp}). For smaller $J_{\perp}$ the empirical fit is not good due to finite size effects in our calculations . Taking into account the complication of performing quantum Monte Carlo (QMC) calculations at very low temperatures on triangular lattices --due to the sign problem-- the fact that the best fit gives a value for the spin stiffness that is close to the SP result already computed at zero temperature, $\rho _s=0.11$~\cite{Manuel1998}, lends support to our procedure to estimate $J_\perp$ values from $T_N$. We remark that, within the SBT, a slightly different formula for $T_N$ as a function of $J_\perp$ has been obtained~\cite{Keimer1992}, but is valid for very small $J_\perp$ (in the region where our calculations suffer from size effects). \\

\begin{figure*}[!t]
\vspace*{0.cm}
\includegraphics*[width=0.9\textwidth,angle=0]{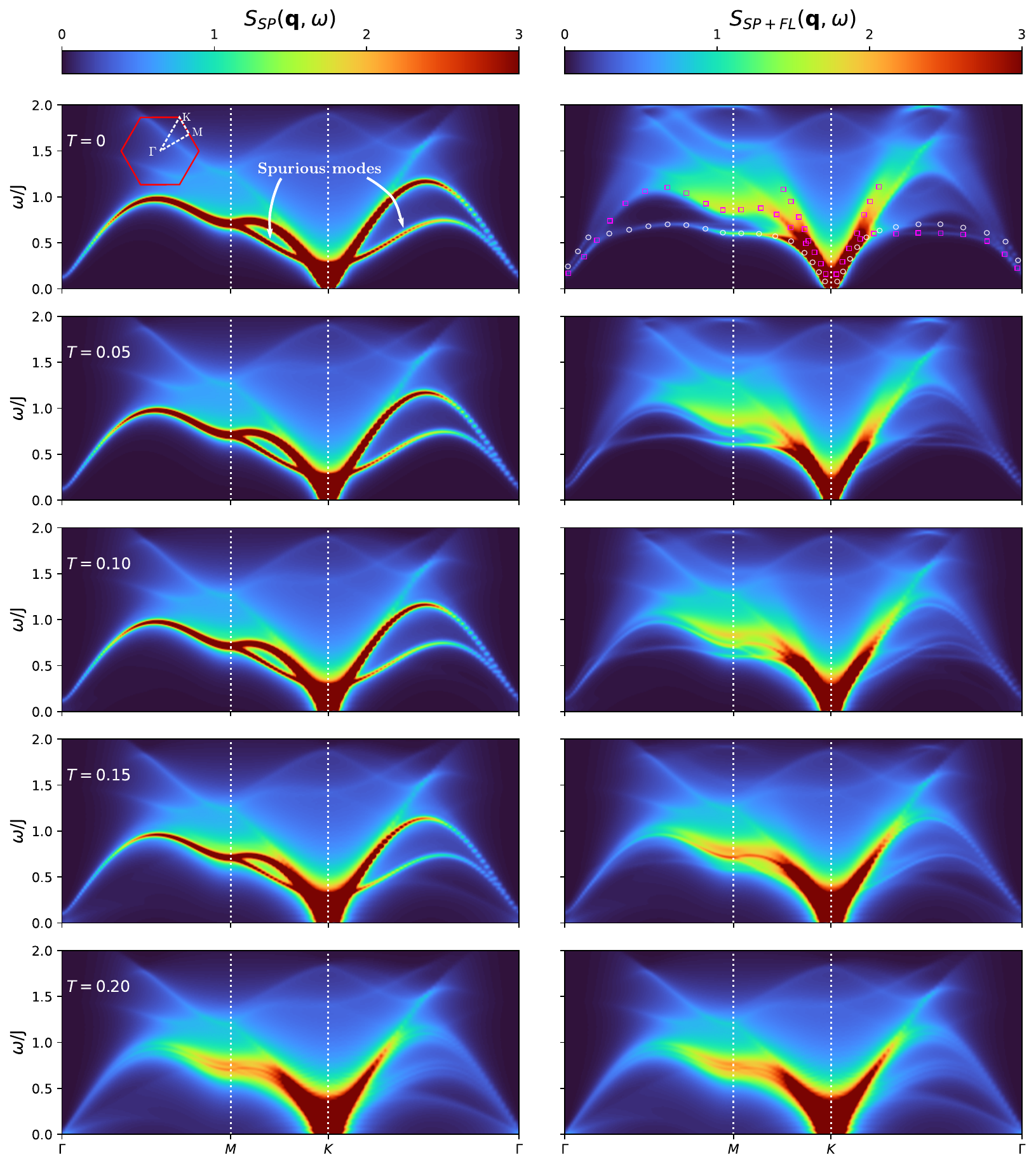}
\caption{Total dynamical spin structure factor (DSSF) for the quasi-2D triangular antiferromagnet ($J_\perp = 0.01J$) obtained with the Schwinger boson theory at different temperatures, along the Brillouin zone path indicated in the top left panel. Left column: saddle point  $S_{SP}(\bm q,\omega)$. Right column: saddle point plus Gaussian fluctuation $S_{SP+FL}(\bm q,\omega)$. At the Néel temperature $T = 0.2 J$, Gaussian fluctuation contributions vanish and the remaining signal comes from the saddle point. Notice that there is a small gap at $\Gamma$ as the spectra are computed for $k_z=\pi$. The open symbols at the top right panel aid to identify the two-spinon bound state bands (poles of the fluctuation matrix): the white circles 
and the magenta squares signal the poles corresponding mainly to out-of-plane $S^{zz}$  and in-plane $S^{xx}+S^{yy}$, respectively. The white arrows in the top left panel signal the SP spurious modes (see main text). The artificial broadening is taken $\eta=0.02$.}
\label{fig:spectrotot}
\end{figure*}

The above results for the Heisenberg model can be easily extended to the $XXZ$ and $J_1-J_2$ models. In both cases we have observed that T$_N$ behaves in the expected manner, that is, for a fixed value of $J_\perp$, T$_{N}$ increases as anisotropy $\Delta$ increases; whereas T$_{N}$ decreases as frustration $J_2/J_1$ increases. In Ba$_3$CoSb$_2$O$_9$, described by the $XXZ$ model with estimated values $J^z/J=0.937$ and $J_\perp / J= 0.061$, the N\'eel temperature gives $T^{\rm XXZ}_{N}/J =  0.274$, which should be compared with the experimental value $T^{\rm exp}_N/J^{\rm exp} \sim 0.2$ \cite{Ito2017,Macdougal20,Ghioldi2022}. In KYbSe$_2$, described by the $J_1-J_2$ model with estimated value $J_2/J_1=0.05$ and the N\'eel temperature $T^{\rm exp}_N / J_1 ^{\rm exp} \sim 0.045$ \cite{Scheie24} we obtain a magnitude for $J_{\perp}$ that is smaller than $0.01\%$ of $J$. 
This result is in line with the experimental certainty that KYbSe$_2$ is one of the best examples of a two dimensional triangular AF. 

Finally, it is important to remark that although we do not expect to obtain a precise prediction for the values of $J_{\perp}$ or $T_N$ from a saddle point (mean field) approximation, the fact that the SP solution gives the expected qualitative description of finite temperature properties is crucial to compute reliable spectra (to order $1/\mathcal{N}$) in order to get a consistent thermal picture of the magnetic excitations. 

\section{Spectra}

\subsection{Temperature evolution} 

It is well known that the Schwinger boson mean field theory (equivalent to the SP approximation) provides a reliable description of the static properties of quantum antiferromagnets~\cite{Auerbach1994,Ceccatto1993}. However, the approximation fails to qualitatively account for the dynamical properties. Particularly, the true magnon excitations for an ordered state are not captured by the saddle-point approximation~\cite{Chandra1991,Ghioldi2018,Zhang2019,Zhang2021}. It is then necessary to go beyond the SP level, including Gaussian corrections to properly account for the magnons. Given the proximity of the quasi-2D triangular antiferromagnet to a quantum phase transition, its dynamical properties exhibit  mixed features from the close spin liquid phase (fractionalized excitations, associated with a continuum spectra) and the long-range magnetic order (magnons). This coexistence is properly accounted for by our $1/\mathcal{N}$- Schwinger boson approach, where the magnetic excitations of the quasi-2D triangular antiferromagnet are better described in terms of fractionalized excitations (spinons), and magnons that are recovered as two-spinon bound states. This hypothesis has been corroborated in previous works
\cite{Ghioldi2022,Scheie24}.

In this section we focus on the temperature evolution of the magnetic spectra of the quasi-2D triangular antiferromagnet. Namely, we move along the red vertical line of Fig. \ref{fig1:diagram}. For this purpose, we compute the dynamical spin structure factor (DSSF) (see Eq.~\ref{eq:dssf}), taking $J_\perp/J = 
10^{-2}$. The saddle-point DSSFs are displayed in the left column of Fig.~\ref{fig:spectrotot}, along the Brillouin zone path shown in the same figure, while the Gaussian-corrected DSSFs are in the right column.

We begin discussing in detail the zero temperature case, shown in the top panels of Fig.~\ref{fig:spectrotot}. 
At the saddle point level the system is described as a gas of independent spinons \cite{Mezio2011} with dispersion relation $\varepsilon^\sigma_{\bm k}$. It is worthwhile to stress that $\varepsilon^\sigma_{\bm k}$ should not be identified as the magnon excitation of the spin-wave theory \footnote{Incidentally, for the square lattice, the SBMFT $\varepsilon^\sigma_{\bm k}$ closely resemble LSWT dispersion relation. However, this is not the case for non-collinear magnetic orderings, like the $120^\circ$ Néel order in the triangular lattice.}. Actually, within the SBT, any physical spin-1 excitation is composed of two spinons, and they must be extracted from the DSSF. As the spinons are independent at SP level, the DSSF shows a two-spinon continuum (branch cut), in contrast to the predictions of linear spin wave theory, 
where magnons are well defined as $\delta$-peaks. The top left panel of Fig.~\ref{fig:spectrotot} shows that the saddle point continuum has an intense lower edge, which for momentum $\textbf{q}$ corresponds to $\min_\textbf{k}\{\varepsilon^{\sigma'}_{\bm k+\bm q}+\varepsilon^\sigma_{\bm k}\}$ (see 
denominator in Eq.~(\ref{eq:chizzsp})). This minimum is obtained when one spinon excitation is created in the normal phase and another in the condensate with momenta $\textbf{k}^* (= \pm \frac{\bm Q}{2})$. Because of that, the lower edge coincides with the uncondensed mode $\varepsilon^{\sigma }_{\bm k ^* +\bm q}$ or $\varepsilon^{\sigma}_{\bm k* -\bm q }$. This high intensity of the lower edge has often led to the wrong identification of the single spinon excitation with the physical magnon. Furthermore, due to the local constraint violation, the saddle point magnetic spectrum also exhibits spurious modes arising from unphysical density fluctuations~\cite{Mezio2012}, indicated by arrows in the top left panel of Fig.~\ref{fig:spectrotot}.

The Gaussian corrections of the dynamical susceptibility~ \cite{Ghioldi2018,Ghioldi2022}, given by Eq. (\ref{eq:corrections}), drastically change the SP spectrum as shown in the top right panel of Fig.~\ref{fig:spectrotot}: i) it cancels out both, the lower edge of the SP two spinon continuum and the spurious modes and ii) it introduces new collective modes (the poles of the RPA propagator $\mathcal{D}$) consisting of low energy two-spinon bound states, along with a quasi-free two spinon continuum at higher energy. These poles are marked in the figure 
with open circles and squares to aid their identification. The circles signal the poles mainly corresponding to $S^{zz}(\bm q,\omega)$ (out-of-plane transverse fluctuations), 
while the squares signal the poles mainly of $S^{xx}(\bm q.\omega)+S^{yy}(\bm q,\omega)$ 
(in-plane mixed longitudinal and transverse fluctuations). Furthermore, the Gaussian-corrected spectra contain a rather intense structured continua, which extend up to
$\simeq 2J$ (three times the magnon bandwith)~\cite{Ghioldi2018,Ghioldi2022}. 

The two-spinon bound states are collective modes, originated by spinon interactions incorporated by the Gaussian fluctuations of the Hubbard-Stratonovich fields. 
These bound states are indeed the single-magnon excitations since, in the large-$S$ limit, we have numerically found that the $1/\mathcal{N}$ SBT spectrum coincides with the LSWT prediction \cite{Zhang2019}. Namely, in the large-$S$ limit, the two-spinon continuum vanishes and the two-spinon bound state bands coincide exactly with both the magnon dispersion relation and the spectral weight predicted by LSWT. Then, for the physical case $S=1/2$, it is fair to identify the two-spinon bound state with the single-magnon excitations. 
It can be seen that there are poles thar are well separated from the continua --for example, the poles corresponding mainly to $S^{zz}(\bm q,\omega)$ (white circles)--, while others are located within the continua. The former are long-lived magnon excitations, while the latter can be thought of as quantum resonances of two-spinons, magnons with a finite lifetime. 
Based on previous works, one can conclude that, at zero temperature, the DSSF features for the quasi-2D triangular AF are similar to those described for the pure two-dimensional model ~\cite{Mezio2011,Ghioldi2018,Zhang2019}.

 The temperature evolution of the magnetic spectra is displayed on Fig.~\ref{fig:spectrotot}. 
On the left column, it can be seen that there is almost no change in the structure and intensity of the saddle point spectra, except very close to the Néel temperature (bottom left panel). In contrast, the $1/\mathcal{N}$-corrected results, shown in the right column, exhibit a non-trivial behavior with temperature. As temperature increases, a continuum signal starts filling the gap between the lower single-magnon dispersions and the $T=0$ continuum, due to both thermal processes:  simultaneous creation and absorption of spinons (see denominators $\varepsilon^{\sigma'}_{\bm k + \bm q}-\varepsilon^{\sigma}_{\bm k}$ in Eq.~(\ref{eq:pifl})) and creation of two spinons
 (see denominators $\varepsilon^{\sigma'}_{\bm k + \bm q}+\varepsilon^{\sigma}_{\bm k}$ ).   
Processes of the first kind take place at the same energy of the lower magnon bands, destroying the coherence of the two-spinon bound states. So, the magnons decay into two quasi-free spinons, restoring part of the signal of the saddle-point lower edge. Consequently, the single-magnon excitations below the $T=0$ continua acquire a finite lifetime, signalled by the broadening of the magnon quasiparticle (see Subsection C). In a semiclassical picture, at $T=0$, due to the non-collinear magnetic order,  the transverse and longitudinal magnetic fluctuations are tightly coupled. Then, the effect of finite temperature is to incoherently decouple both kind of fluctuations, weakening the magnon excitations. The $1/\mathcal{N}$ corrections correctly capture this behavior.

\begin{figure}[!th]
\vspace*{0.cm}
\includegraphics*[width=0.5\textwidth,angle=0]{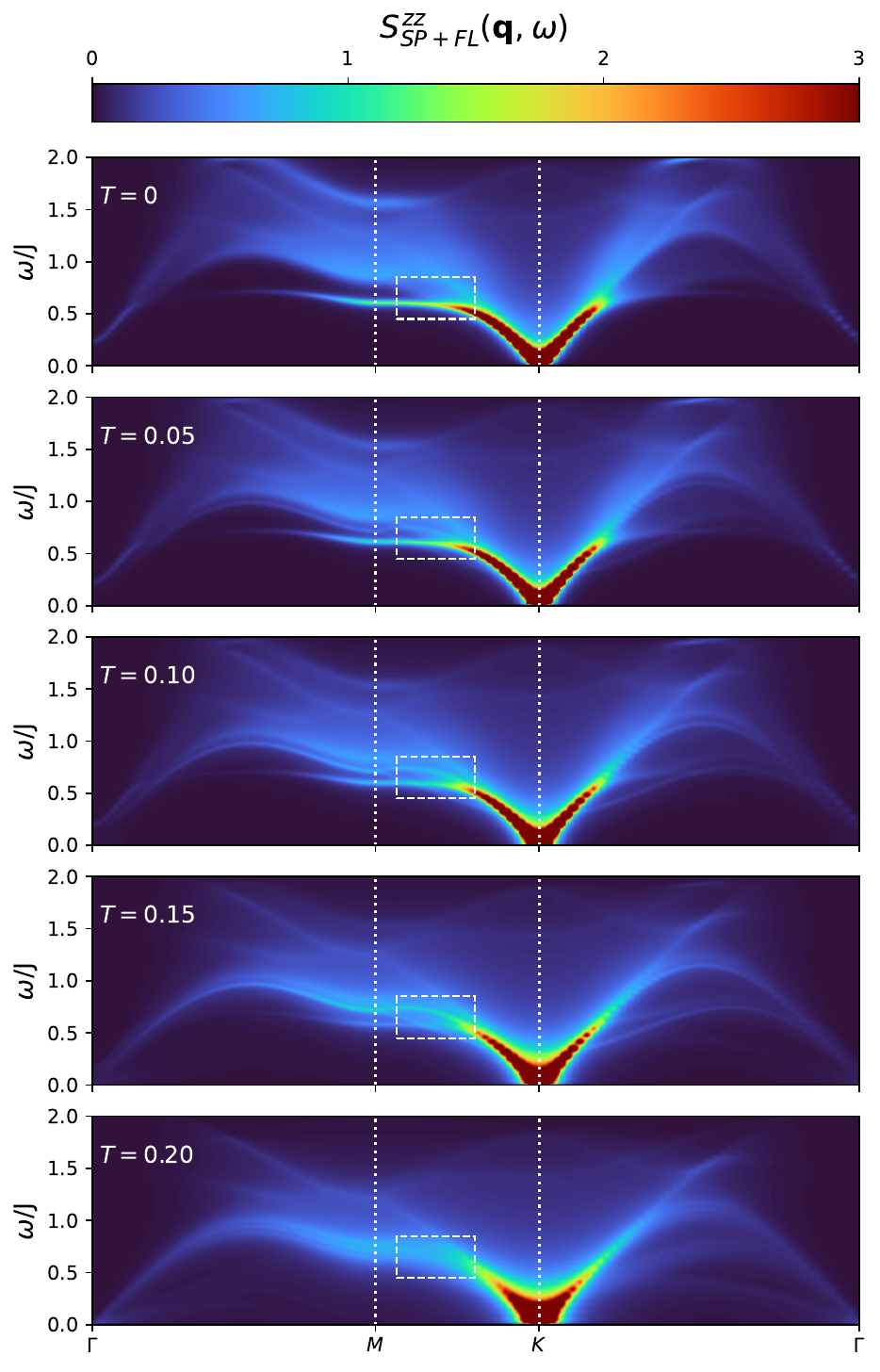}
\caption{Thermal evolution of the transverse dynamical structure factor $S^{zz}(\textbf{q},\omega)$ for the quasi 2D triangular antiferromagnet ($J_\perp = 0.01J$) obtained with the Schwinger boson theory. Inside the dashed rectangle it is shown how the emergent thermal continuum mixes with the magnon band at $T=0.15J$, producing its decay.}
\label{fig2:espectro2}
\end{figure}

\begin{figure}[!ht]
\vspace*{0.cm}
\includegraphics*[width=0.5\textwidth,angle=0]{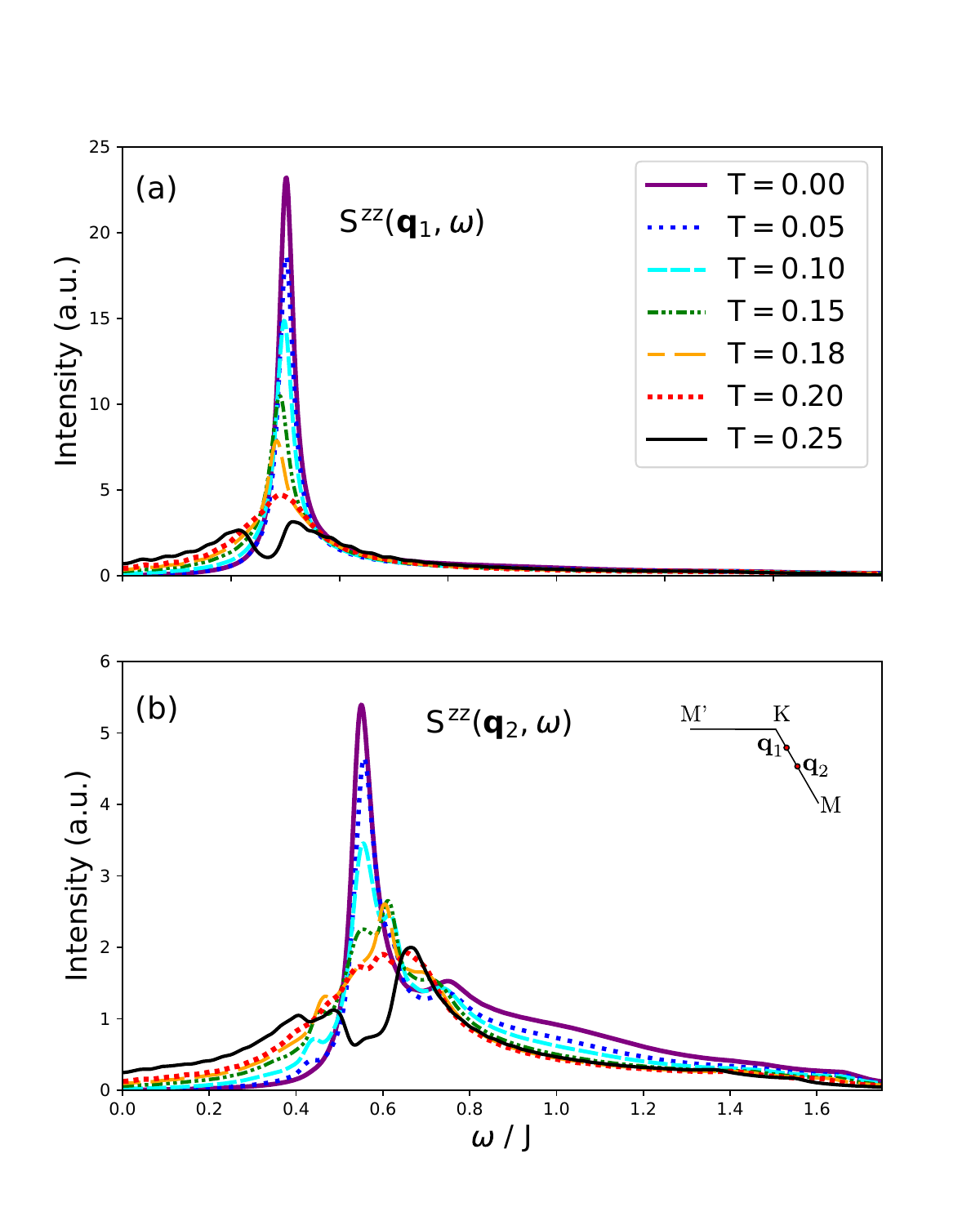}
\caption{Thermal evolution of the transverse dynamical structure factor $S^{zz}(\textbf{q},\omega)$ for (a) $\textbf{q}_1\! = \!( \frac{3\pi}{4} , \frac{7\pi}{4\sqrt{3}},\pi )$ and (b) $\textbf{q}_2\!=\!( \frac{5\pi}{6} , \frac{3\pi}{2\sqrt{3}},\pi )$. The two momenta are shown in a plot of the BZ path at the top right corner of the (b) panel. The magnon peak signal holds to higher temperatures for the closest momentum to the Goldstone mode, $\textbf{q}_1$. At the Néel temperature $T \!=\!0.2 J$, the fluctuations vanish completely, and the excitations are only due to free spinons.}
\label{fig:corte}
\end{figure}

In what follows, we concentrate on the $zz$ component of the dynamical spin structure factor, because it  better displays long-lived magnon excitations.
Fig.~\ref{fig2:espectro2} shows the Gaussian-corrected $S^{zz}(\bm q,\omega)$  along the same path of the Brillouin zone as Fig.~\ref{fig:spectrotot}. As the long-range magnetic order lies in the $x-y$ plane, $S^{zz}(\bm q, \omega)$ only takes into account transverse magnetic fluctuations, whereas the total spectra also contain the longitudinal fluctuations, that for non-collinear orders are tightly coupled with the in-plane transverse ones. It can be noticed that the relative spectral weight of the continuum signal with respect to the single-magnon signal is smaller for the $zz$ component than for the total one (see right column in Fig.~\ref{fig:spectrotot}). In other words, the anomalous extended continua, characteristic of the triangular quantum antiferromagnet, has its origin mainly in the in-plane spin fluctuations.  The white dashed rectangle is a visual aid to help follow the thermal evolution of a sector of the single-magnon dispersion, between the $K$ and $M$ points. For low temperatures, the magnon excitations in this sector are well defined as there is almost no change with respect to $T=0$. However, around $T \simeq 0.15J,$ the thermally activated spinons produce the decay of the two-spinon bound states (magnons).  Beyond this temperature, the whole spectra become diffusive except in the neighborhood of the Goldstone modes. 
This behavior is also illustrated  in Fig.~\ref{fig:corte} which displays $S^{zz}(\bm q,\omega)$ as a function of energy for two different momenta (see Subsection IV B).

Concerning the magnon decay, the spectra suggest the existence of a crossover temperature, above which the magnon excitations lose their coherence across a significant region of the first Brillouin zone, due to the mentioned decay into the emergent continuum.  Only in the vicinity of the Goldstone modes, magnons survive as well-defined quasiparticles. Ultimately, above the Néel temperature, the spectra become fully diffusive, and the $1/\mathcal{N}$ corrections to the magnetic dynamical susceptibility vanish. Notice that saddle-point and $1/\mathcal{N}$-corrected results are practically the same (see $T=0.20 J$ panel in Fig.~\ref{fig:spectrotot}).  This is equivalent to saying that, when the temperature is high enough, the $SU(2)$ symmetry is restored.

\subsection{Relative spectral weight of magnons and the crossover to the terminated Goldstone regime} 
\begin{figure}[!ht]
\vspace*{0.cm}
\includegraphics*[width=0.48\textwidth,angle=0]{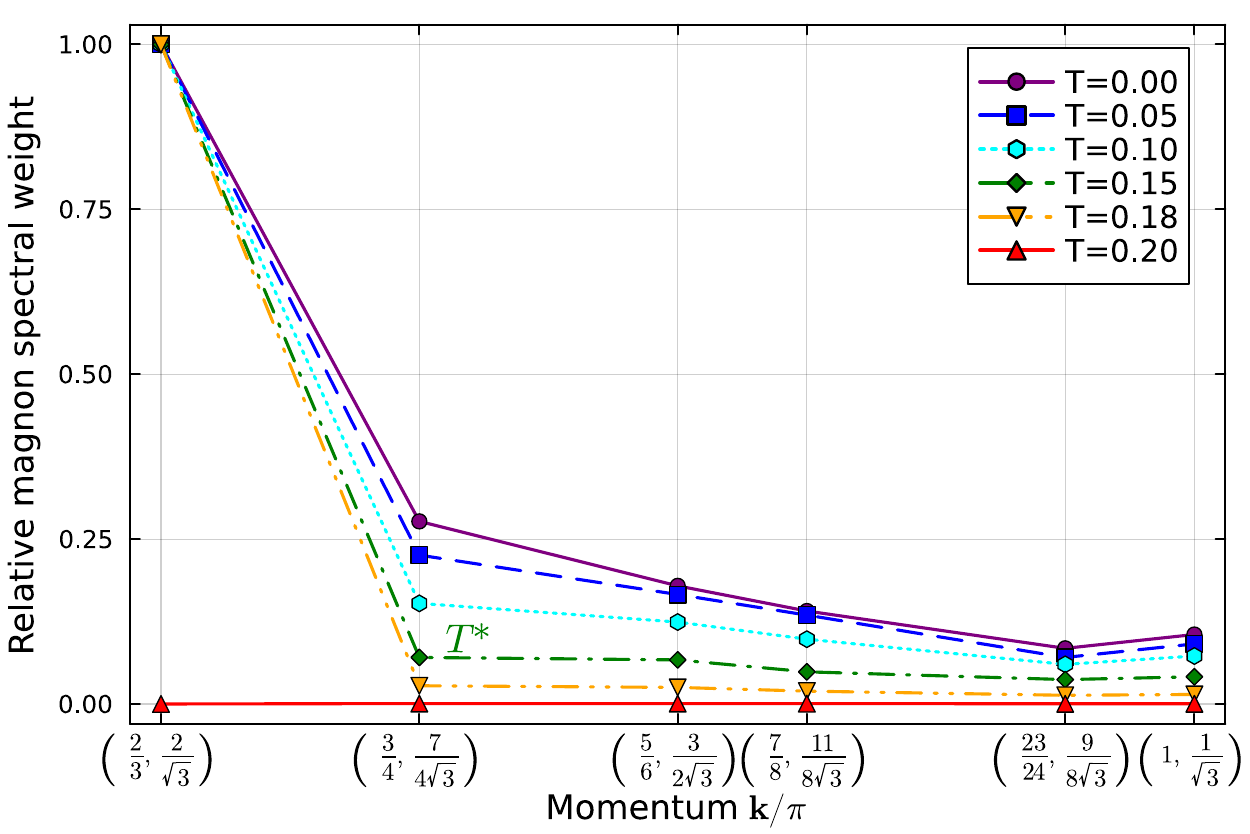}
\caption{Relative magnon spectral weight as a function of the temperature for some momenta $\textbf{k}$ on the Brillouin zone path $K = 
 (\frac{2\pi}{3},\frac{2\pi}{\sqrt{3}},\pi) \longrightarrow M = (\pi, \pi/\sqrt{3},\pi)$. 
On the x-axis, the component common to all points, $k_z = \pi$, is omitted.}
\label{fig:weight}
\end{figure}

To support our picture about the thermal evolution of the magnons and to identify a crossover temperature $T^*$ (along the red vertical line of Fig.~\ref{fig1:diagram}), we analyze the spectral weight contribution of the two-spinon bound states and fix a criterion to quantify $T^*$. 
Fig.~\ref{fig:corte} displays $S^{zz}(\bm q,\omega)$ for two different momenta $\bm q_1$ and $\bm q_2$. 
It can be seen that for $\bm q_1$, close to Goldstone point $K$, although the magnon loses spectral weight as the temperature increases, it is always well defined, up to the Néel temperature ($T_N = 0.2 J$). On the other hand, for $\bm q_2$ in the middle between the $K$ and $M$ points, the quasiparticle coherence is lost for $T \gtrsim 0.15 J$, being its spectral weight mostly transferred to the continuum background.  This behaviour is observed in an ample region of the Brillouin zone. In order to see this, in Fig.~\ref{fig:weight} we  compare the magnon spectral weight with the total signal of the dynamical structure factor along the $K \to M$ path, for several temperatures. 
The points are computed by taking the quotient between the spectral weight of the magnon resonance,  
obtained from the $1/\mathcal{N}$-corrected $S^{zz}(\bm q,\omega)$, and the total (saddle-point plus $1/\mathcal{N}$-corrections) spectral weight. We identify the crossover temperature $T^* \simeq$ 0.75$T_N$ ($T^* \simeq 0.15 J$ for $J_\perp/J = 10^{-2})$, defined as the one above which, for all momenta except in the vicinity of the Goldstone modes, the magnon spectral weight goes down to 10$\%$ of the total spectral weight. 
The vicinity of the Goldstone modes is taken as a region around them with a radius that is $\simeq 25 \%$ of the $K-M$ distance. This criterion is consistent with the results shown in Fig.~\ref{fig:corte}.
A similar behavior is observed in the excitation spectrum of the superfluid $^4$He, where phonon excitations terminate at some momentum by decaying into two roton excitations \cite{Stone2006,Zhitomirsky2013}. Though the decay mechanism differs, we term this antiferromagnetic ordered region, between $T^{*}$ and $T_{N}$, the \textit{terminated Goldstone regime} (see Fig.~\ref{fig1:diagram}). This observation contrasts with the expectation that the magnon would persist throughout the entire Brillouin zone for non-frustrated antiferromagnets. Finally, at the Néel temperature, the whole spectra become incoherent, even at the Goldstone modes.

\subsection{Lifetime of the magnon excitations}

\begin{figure}[ht]
\vspace*{0.cm}
\includegraphics*[width=0.48\textwidth,angle=0]{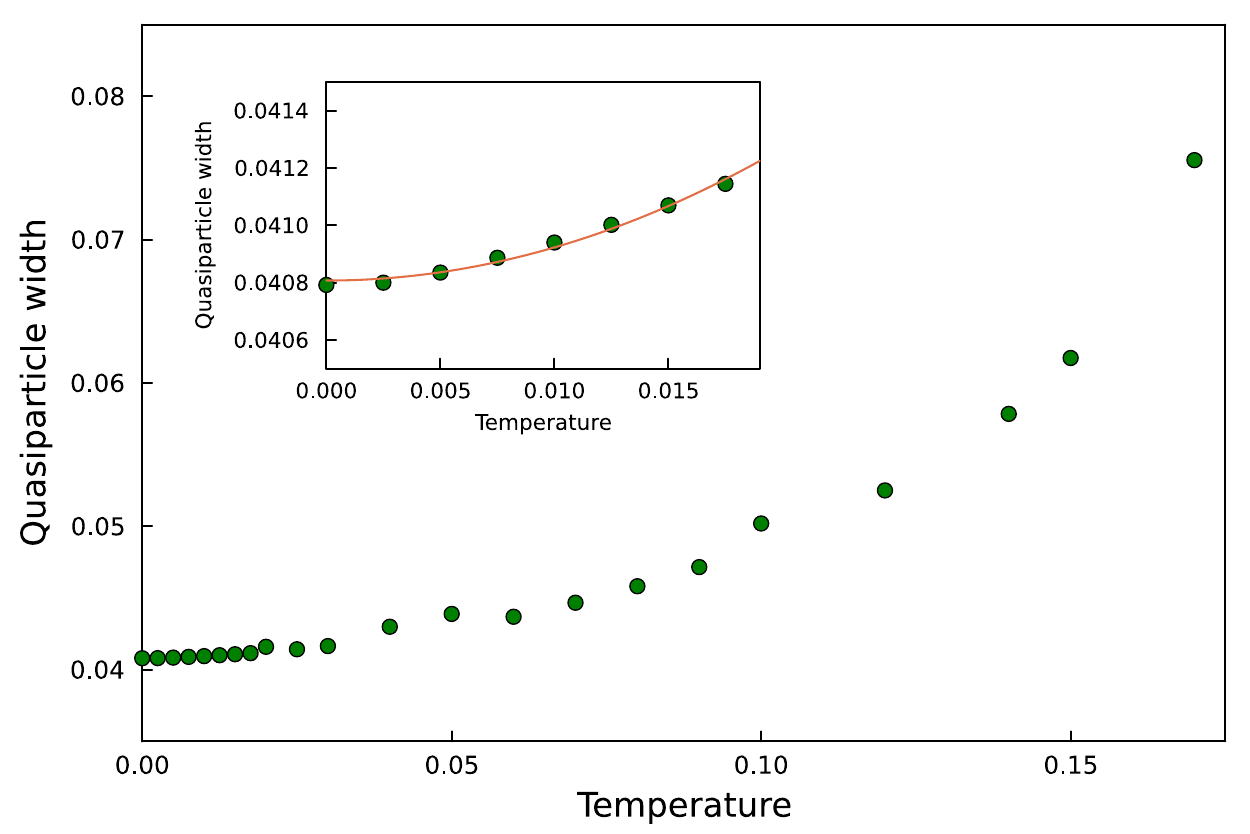}
\caption{Temperature dependence of the quasiparticle peak width for momentum $\textbf{k}=(2.225,3.4,\pi)$  which is close the Goldstone mode. Inset: Low temperature fit of the form $f(T) = aT^2+b$, where $a=1.1555$ and $b=0.0408$. $b$ is twice the artificial broadening $\eta$.}
\label{width}
\end{figure}

In order to further quantify the effect of the thermal activated spinons on the two-spinon bound states,  Fig.~\ref{width} shows the temperature evolution of its width for a momentum $\bm k$ close to the Goldstone mode. While at $T=0$ the finite width is given by twice the artificial broadening $\eta = 0.02$, necessary to perform the analytical continuation (see Eq.~(\ref{eq:dssf})), as the temperature rises the width has a $T^2$ behavior (see inset). At low temperature ($T < T^*$) the lifetime hardly changes with respect to its ground state value whereas for $T > T^*$ there is a pronounced increase. This is consistent with the terminated Goldstone regime picture. It is important to mention that the thermal evolution of the magnon lifetime has been little explored in the literature~\cite{Bayrakci13}. We recall that the present result is limited to two-spinon scattering processes, and if we would like to include magnon-magnon interactions, four spinon processes should be considered~\cite{Ghioldi2018}.

\section{Concluding remarks}
Our results on the quasi-2D triangular antiferromagnet can be related with recent numerical predictions for the specific heat and entropy of the 2D triangular   Heisenberg antiferromagnet~\cite{Prelovsek18,Chen19,Gonzalez22}. In these studies two characteristics temperatures are found: a high temperature scale $T_h \sim 0.5 J$ signaling the development of nearest-neighbor spin correlation and a low temperature scale $T_l \sim 0.2 J$ related to the thermal activation of the roton excitations. We conjecture that the crossover temperature $T^* \simeq 0.75 T_N$ can be associated with $T_l:$ 
around $T \sim T^*$ the antiferromagnet enters the terminated Goldstone regime, where the high density of thermally activated spinons around $M$ gives rise to the decay of the magnons away from the Goldstone points. This suggests that the change of the nature of the magnetic excitations around $T^*$ notoriously affects the thermodynamics of the triangular quantum antiferromagnet. 
On the other hand, the high temperature scale $T_h$ can be associated with the "pathological" temperature $T \simeq 0.43 J$ of the Schwinger boson theory~\cite{Tchernyshyov2002,Zhang2002}, above which the mean-field solution corresponds to an infinite temperature perfect paramagnet, in which the spin correlations vanish exactly. 

In summary, in order to make contact with the experimental conditions we have studied the thermal evolution of the magnetic excitations of a quasi-2D triangular antiferromagnet, by means of a parton theory solved as a $1/\mathcal{N}$ expansion of Schwinger boson theory. Recently, this approach has been successfully applied --at zero temperature-- to frustrated antiferromagnets close to a quantum phase transition~\cite{Ghioldi2022,Scheie24}. We have found that the magnon excitations, described in our theory as two-spinon bound states, decay due to the proliferation of thermally activated spinons.  The decay is very pronounced in the temperature range $T^* < T < T_N$, which we have named {\it terminated Goldstone regime}, because --above $T^*$-- the two-spinon bound states survive only in the vicinity of the Goldstone modes.  In the quantum critical regime, above the Néel temperature, the excitations would correspond to quasi-free spinon states.

The coexistence of quantum critical features and magnon excitations, below Néel temperature, has  been observed in weakly 3D interacting chains~\cite{Lake2005} and also in the triangular antiferromagnet CuCrO$_2$~\cite{Kajimoto2015}. These results gives support to the idea that the fractionalization of magnons --proposed for 2D frustrated AFs-- near a quantum phase transition~\cite{Chubukov1994,Chubukov1995,Sachdev2008,Zhou2017} can be safely extended to more realistic quasi-2D triangular antiferromagnets.

\section{Acknowledgments}
We thank Cristian Batista for insightful discussions.  
We acknowledge financial support by CONICET (Argentina) under Grant PIP No. 3220.
\appendix

\section{Numerical inverse of the dynamical matrix}
The dynamical matrix (\ref{invgreensp}) can be written as
\begin{equation}
    \mathcal{G}_{sp} ^{-1} (\textbf{k},i\omega) = i \omega \mathcal{S} + \mathcal{M}_\textbf{k}
\end{equation}
where we have separated the $\textbf{k}$ dependent terms from the $\omega$ dependent ones, and $\mathcal{S}=diag(+1,-1,+1,-1)$ is the paraunitary matrix. Then we perform a paraunitary diagonalization of the momentum dependent part, $\mathcal{M}_\textbf{k} = \mathcal{U}^\dagger _\textbf{k} \mathcal{D}_\textbf{k} \mathcal{U} _\textbf{k}$, where $\mathcal{D}_\textbf{k}$ is the diagonal matrix that contains the eigenvalues and $\mathcal{U}_\textbf{k}$ is the paraunitary transformation matrix \cite{Colpa1978}. Given that $\mathcal{S}$ is invariant under paraunitary transformations ($S = \mathcal{U}^\dagger _\textbf{k} S \mathcal{U} _\textbf{k}$) we can write
\begin{equation}
    \mathcal{G}_{sp} ^{-1} (\textbf{k},i\omega) = \mathcal{U}^\dagger _\textbf{k}
 \left(i \omega \mathcal{S} + \mathcal{D}_\textbf{k} \right) \mathcal{U} _\textbf{k}
\end{equation}
and the dynamical matrix inverse is obtained as
\begin{equation}
    \mathcal{G}_{sp}  (\textbf{k},i\omega) = \mathcal{U}^{-1} _\textbf{k}
     \mathcal{S}
 \left(i \omega +  \mathcal{S}\mathcal{D}_\textbf{k} \right)^{-1} \left( \mathcal{U}^\dagger _\textbf{k} \right) ^{-1}
\end{equation}
where
\begin{equation}
    \left(i \omega+  \mathcal{S} \mathcal{D}_\textbf{k} \right)^{-1} = \begin{pmatrix}
        \frac{1}{i\omega + \varepsilon^+ _\textbf{k}} & 0 & 0 & 0 \\
        0 &  \frac{1}{i\omega - \varepsilon^- _\textbf{k}} & 0 & 0 \\
        0 & 0 &  \frac{1}{i\omega + \varepsilon^- _\textbf{k}} & 0 \\
        0 & 0 & 0 &  \frac{1}{i\omega - \varepsilon^+ _\textbf{k}}
    \end{pmatrix}.
\end{equation}
The eigenvalues $\varepsilon^\sigma _\textbf{k}$ structure is consistent with the $\textbf{k} \rightarrow -\textbf{k}$ invariance.
Finally, the simple fraction decomposition of the SP Green function matrix is
\begin{equation}
   \mathcal{G}_{sp}  (\textbf{k},i\omega) =  \sum _{\sigma \sigma ' }\frac{g^{\sigma \sigma '} (\textbf{k})}{i\omega + \sigma \varepsilon^{\sigma '} _\textbf{k}}  
\end{equation}
where $\sigma , \sigma ' = \pm$ and
\begin{equation*}
    \begin{split}
            g ^{++} _{\alpha \beta} = \mathcal{U}^{-1} _{\alpha 1} \left( \mathcal{U}^{\dagger} \right) _{1\beta} ^{-1}, \hspace{0.5cm}  g ^{--}  _{\alpha \beta} = -\mathcal{U}^{-1} _{\alpha 2} \left( \mathcal{U}^{\dagger} \right) _{2\beta} ^{-1}, \\
         g ^{+-} _{\alpha \beta} = \mathcal{U}^{-1} _{\alpha 3} \left( \mathcal{U}^{\dagger} \right) _{3\beta} ^{-1}, \hspace{0.5cm}   g ^{-+}  _{\alpha \beta} = -\mathcal{U}^{-1} _{\alpha 4} \left( \mathcal{U}^{\dagger} \right) _{4\beta} ^{-1}.
    \end{split}
\end{equation*}
In the last expressions we have omitted the $\textbf{k}$ dependence to simplify the notation.

   \bibliography{version_final}

\end{document}